# Approximate relativistic solutions for a new ring-shaped Hulthén potential

Sameer M. Ikhdair[1*], Majid Hamzavi[2**]

[1] *Physics Department, Near East University, 922022 Nicosia, North Cyprus, Turkey*

*and*

[1] *Physics Department, Faculty of Science, An-Najah National University, Nablus, West Bank, Palestine*

[2] *Department of Basic Sciences, Shahrood Branch, Islamic Azad University, Shahrood, Iran*

[*] *sikhdair@neu.edu.tr*

*Tel: +90-392-2236624; Fax: +90-392-2236622*

[**] Corresponding Author: *majid.hamzavi@gmail.com*

*Tel: +98-273-3395270; Fax: +98-273-3395270*

**Abstract**

Approximate bound state solutions of the Dirac equation with the Hulthén plus a new generalized ring-shaped (RS) potential are obtained for any arbitrary $l$-state. The energy eigenvalue equation and the corresponding two-component wave function are calculated by solving the radial and angular wave equations within a recently introduced shortcut of Nikiforov-Uvarov (NU) method. The solutions of the radial and polar angular parts of the wave function are given in terms of the Jacobi polynomials. We use an exponential approximation in terms of the Hulthén potential parameters to deal with the strong singular centrifugal potential term $l(l+1)r^{-2}$. Under the limiting case, the solution can be easily reduced to the solution of the Schrödinger equation with a new ring-shaped Hulthén potential.

# 1- Introduction

The spin or pseudospin (p-spin) symmetry [1,2] investigated in the framework of the Dirac equation is one of the most interesting phenomena in the relativistic quantum mechanics to explain different aspects for nucleon spectrum in nuclei. This is mainly studied for the existence of identical bands in superdeformed nuclei in the framework of a Dirac Hamiltonian with attractive scalar $S(r)$ and repulsive vector $V(r)$ potentials [3]. The p-spin symmetry is based on the small energy difference between single-nucleon doublets with different quantum numbers and the Hamiltonian of nucleons moving in the relativistic mean field produced by the interactions between nucleons. The relativistic dynamics are described by using the Dirac equation only [4].

Recently, Ginocchio [5] found that the p-spin symmetry concept in nuclei occurs when $S(r)$ and $V(r)$ potentials are nearly equal to each other in magnitude but opposite in sign, i.e., $S(r) \approx V(r)$ and hence their sum is a constant, i.e., $\Sigma(r) = S(r) + V(r) = C_{ps}$. A necessary condition for occurrence of the p-spin symmetry in nuclei is to consider the case $\Sigma(r) = 0$ [5-7]. Further, Meng et al [8] showed that the p-spin symmetry is exact under the condition of $d\Sigma(r)/dr = 0$. Lisboa et al [9] studied the generalized harmonic oscillator for spin-1/2 particles under the condition $\Sigma(r) = 0$ or $\Delta(r) = S(r) - V(r) = 0$. The Dirac equation has been solved numerically [10,11] and analytically [4,12,13] for nucleons that are moving independently in the relativistic mean field in the presence of the p-spin symmetric scalar and vector potentials. Thus, the exact or approximate analytical solution of the Dirac equation leads to the bound-state energy spectra and spinor wave functions [14,15].

Over the past years the study of exact and approximate solutions of the Schrödinger, Klein-Gordon (KG) and Dirac wave equations with non-central potentials becomes of considerable interest. For example, a new ring-shaped (RS) potential has been introduced [16] plus Coulomb potential [16], Hulthén potential [17], modified Kratzer potential [18] and non-harmonic oscillator potential [19]. Such calculations with this RS potential have found applications in quantum chemistry such as the study of ring-shaped molecules like benzene. Furthermore, the shape forms of this potential play an important role when studying the structure of deformed nuclei or the nuclear interactions. Quesne [20] obtained a new RS potential by replacing the Coulomb part of Hartmann potential [21] by a harmonic oscillator term. Chen [22] exactly obtained energy spectrum of some non-central separable potential in $r$ and $\theta$ using the method of supersymmetric WKB approximation. Yaşuk *et al* [23] obtained general solutions of Schrödinger equation for a non-central potential by using the NU method [24]. Chen *et al.* [25] studied exact solutions of scattering states of the KG equation with Coulomb potential plus new RS potential with equal mixture of scalar and vector potentials. Ikhdair and Sever [26] used the polynomial solution to solve a non-central potential. Zhang and Wang [27] studied the KG with equal scalar and vector Makarov potentials by the factorization method. Kerimov [28] studied non-relativistic quantum scattering problem for a non-central potential which belongs to a class of potentials exhibiting an accidental degeneracy. Berkdemir and Sever [29] investigated the diatomic molecules subject to central potential plus RS potential. Berkdemir and Sever [30] studied the p-spin symmetric solution for spin-$1/2$ particles moving under the effect of the Kratzer potential plus an angle-dependent potential. Yeşiltaş [31] showed that a wide class of non-central potentials can be analyzed via the improved picture of the NU method. Berkdemir and Cheng [32] investigated the problem of relativistic motion of a spin-$1/2$ particle in an exactly solvable potential consisting of harmonic oscillator potential plus a novel RS dependent potential. Zhang *et al* [33-35] obtained the complete solutions of the Schrödinger and Dirac equations with a spherically harmonic oscillatory RS potential. Ikhdair and Sever obtained the exact solutions of the D-dimensional Schrödinger equation with RS pseudo-harmonic potential [36], modified Kratzer potential [37] and the D-dimensional KG equation with ring-shaped pseudo-harmonic potential [38]. Hamzavi *et al* found the exact solutions of Dirac equation with Hartmann potential [39] and RS pseudo-harmonic oscillatory potential [40] by using NU method. Many authors have also studied a few non central potentials within the supersymmetric quantum mechanics and point canonical transformations [41-43].

The aim of this work is to investigate analytically the bound-state solutions of the Dirac equation with non central Hulthén potential plus a new generalized RS potential with extra additional parameter $\alpha$ from the RS potential being used in [21]. Therefore, the non central potential of the type $V(\vec{r}) = V_H(r) + \frac{1}{r^2} V_{RS}(\theta)$, consisting of two parts:

$$V_H(r) = -\frac{V_0}{e^{\delta r} - 1}, \quad V_{RS}(\theta) = \frac{\alpha + \beta \cos^2 \theta}{\sin^2 \theta}, \quad \delta = \frac{1}{a}, \tag{1}$$

with $V_H(r)$ is the Hulthén potential in which $V_0$, $\delta$ and $a$ are the potential depth, the screening parameter and the range of the Hulthén potential, respectively. Further, $V_{RS}(\theta)$ is a new RS potential identical to the RS part of the Hartmann potential [16]. Here $\alpha = -p\sigma^2\eta^2 a_0^2\varepsilon_0$ and $\beta = -p\sigma^2\eta^2 a_0^2\varepsilon_0$, where $a_0 = \frac{\hbar^2}{me^2}$ and $\varepsilon_0 = -\frac{me^4}{2\hbar^2}$ represent the Bohr radius and the ground-state energy of the hydrogen atom, respectively. Further, $\eta$, $\sigma$ and $p$ are three dimensionless parameters. Generally speaking, $\eta$ and $\sigma$ vary from about 1 up to 10 and $p$ is a real parameter and its value is taken as 1. At first to show the shape of the potential (1), we plot it in figure 1 by taking the following parameter values: $V_0 = 0.1\, fm^{-1}$, $\delta = 0.1$, $\alpha = 1$ and $\beta = 10$.

Secondly, in our solution, we will use a powerful shortcut of the NU method [24] that has proven its efficiency and easily handling in the treatment of problems with second-order differential equations of the type $y'' + (\tilde{\tau}/\sigma)y' + (\tilde{\sigma}/\sigma^2)y = 0$ which are usually encountered in physics such as the radial and angular parts of the Schrödinger, KG and Dirac equations [36-40].

This paper is organized as follows. In section 2, we present the Dirac equation for the generalized RS Hulthén potential. Section 3 is devoted to derive the approximate bound-state energy eigenvalue equation and the associated two-components of the wave function consisting from radial and angular parts within a shortcut of the NU method. In Section 4, we present the non-relativistic limits for the RS Hulthén potential and the Hulthén potential. We end with our concluding remarks in Section 5.

**2. Dirac equation for scalar and vector Hulthén plus new ring-shaped potential**

The Dirac equation for a particle of mass $M$ moving in the field of non-central attractive scalar potential $S(\vec{r})$ and repulsive vector potential $V(\vec{r})$ (in the relativistic units $\hbar = c = 1$) takes the form [44]

$$\left[\vec{\alpha}\cdot\vec{p} + \beta(M + S(r)) + V(r)\right]\psi(\vec{r}) = E\psi(\vec{r}), \qquad (2)$$

with $E$ is the relativistic energy of the system and $\vec{p} = -i\vec{\nabla}$ is the three-dimensional (3D) momentum operator. Further, $\vec{\alpha}$ and $\beta$ represent the $4\times 4$ Dirac matrices given by

$$\vec{\alpha} = \begin{pmatrix} 0 & \sigma_i \\ \sigma_i & 0 \end{pmatrix}, \quad \beta = \begin{pmatrix} I & 0 \\ 0 & -I \end{pmatrix}, \qquad i = 1, 2, 3 \tag{3}$$

which are expressed in terms of the three $2 \times 2$ Pauli matrices

$$\sigma_1 = \begin{pmatrix} 0 & 1 \\ 1 & 0 \end{pmatrix}, \quad \sigma_2 = \begin{pmatrix} 0 & -i \\ i & 0 \end{pmatrix}, \quad \sigma_3 = \begin{pmatrix} 1 & 0 \\ 0 & -1 \end{pmatrix}, \tag{4}$$

and $I$ is the $2 \times 2$ unitary matrix. In addition, the Dirac wave function $\psi(\vec{r})$ can be expressed in Pauli-Dirac representation as

$$\psi(\vec{r}) = \begin{pmatrix} \varphi(\vec{r}) \\ \chi(\vec{r}) \end{pmatrix}. \tag{5}$$

Now, inserting Eqs. (3) to (5) into Eq. (2) give

$$\vec{\sigma} \cdot \vec{p}\, \chi(\vec{r}) = (E - M - \Sigma(\vec{r}))\varphi(\vec{r}), \tag{6a}$$

$$\vec{\sigma} \cdot \vec{p}\, \varphi(\vec{r}) = (E + M - \Delta(\vec{r}))\chi(\vec{r}), \tag{6b}$$

where the sum and difference potentials are defined by

$$\Sigma(\vec{r}) = V(\vec{r}) + S(\vec{r}) \text{ and } \Delta(\vec{r}) = V(\vec{r}) - S(\vec{r}), \tag{7}$$

respectively. For a limiting case when $S(\vec{r}) = V(\vec{r})$, then $\Sigma(\vec{r}) = 2V(\vec{r})$ and $\Delta(\vec{r}) = 0$, Eqs. (6a) and (6b) become

$$\vec{\sigma} \cdot \vec{p}\, \chi(\vec{r}) = (E - M - 2V(\vec{r}))\varphi(\vec{r}), \tag{8a}$$

$$\chi(\vec{r}) = \frac{\vec{\sigma} \cdot \vec{p}}{E + M} \varphi(\vec{r}), \tag{8b}$$

respectively, where $E \neq -M$, which means that only the positive energy states do exist for a finite lower-component $\chi(\vec{r})$ of the wave function.

Combining Eq. (8b) into Eq. (8a) and inserting the potential (1), one can obtain

$$\left[ \nabla^2 + E^2 - M^2 + 2(E + M)\left( \frac{V_0 e^{-r/a}}{1 - e^{-r/a}} - \frac{\alpha + \beta \cos^2\theta}{r^2 \sin^2\theta} \right) \right] \varphi_{nlm}(r, \theta, \phi) = 0, \tag{9}$$

where

$$\nabla^2 = \frac{1}{r^2}\left[ \frac{\partial}{\partial r}\left( r^2 \frac{\partial}{\partial r} \right) + \frac{1}{\sin\theta}\frac{\partial}{\partial \theta}\left( \sin\theta \frac{\partial}{\partial \theta} \right) + \frac{1}{\sin^2\theta}\frac{\partial^2}{\partial \phi^2} \right], \tag{10}$$

and

$$\varphi_{nlm}(r, \theta, \phi) = R_{nl}(r) Y_l^m(\theta, \phi),$$

$$R_{nl}(r) = r^{-1} U_{nl}(r),$$

$$Y_l^m(\theta, \phi) = \Theta_l(\theta) \Phi_m(\phi). \tag{11}$$

Inserting Eqs. (10) and (11) into Eq. (9) and making a separation of variables, we finally arrive at the following sets of second-order differential equations:

$$\frac{d^2 U_{nl}(r)}{dr^2} + \left[ E^2 - M^2 - \frac{\lambda}{r^2} + \frac{2(E+M)V_0 e^{-r/a}}{1 - e^{-r/a}} \right] U_{nl}(r) = 0, \tag{12a}$$

$$\frac{d^2 \Theta_l(\theta)}{d\theta^2} + \cot\theta \frac{d\Theta_l(\theta)}{d\theta} + \left[ \lambda - \frac{m^2}{\sin^2\theta} - \frac{2(E+M)(\alpha + \beta\cos^2\theta)}{\sin^2\theta} \right] \Theta_l(\theta) = 0, \tag{12b}$$

$$\frac{d^2 \Phi_m(\phi)}{d\phi^2} + m^2 \Phi_m(\phi) = 0, \tag{12c}$$

where $m^2$ and $\lambda = l(l+1)$ are two separation constants with $l$ is the rotational angular momentum quantum number.

The solution of Eq. (12c) is periodic and must satisfy the periodic boundary condition, that is,

$$\Phi_m(\phi + 2\pi) = \Phi_m(\phi), \tag{13}$$

which in turn gives the solution:

$$\Phi_m(\phi) = \frac{1}{\sqrt{2\pi}} \exp(\pm i m \phi), \quad m = 0, 1, 2, \cdots. \tag{14}$$

The solutions of the radial part (12a) and polar angular part (12b) equations will be studied in the following section.

## 3. Solutions of the radial and polar angular parts

### 3.1. Solution of the angular part

To obtain the energy eigenvalues and wave functions of the polar angular part of Dirac equation (12b), we make an appropriate change of variables, $z = \cos^2\theta$ (or $z = \sin^2\theta$) to reduce it as

$$\Theta_l''(z) + \left[ \frac{(1/2) - (3/2)z}{z(1-z)} \right] \Theta_l'(z) + \frac{1}{z^2(1-z)^2}$$
$$\times \left[ -\frac{1}{4}[\lambda + 2(E+M)\beta]z^2 + [\lambda - m^2 - 2(E+M)\alpha]z \right] \Theta_l(z) = 0, \tag{15}$$

where the boundary conditions require that $\Theta_l(z=0) = 0$ and $\Theta_l(z=1) = 0$. The solution of Eq. (13) can be easily found by using a shortcut of NU method presented in Appendix A. Now, in comparing the above equation with Eq. (A2), we can identify the following constants:

$$c_1 = \frac{1}{2}, \quad c_2 = \frac{3}{2}, \quad c_3 = 1, \quad A = \frac{1}{4}[\lambda + 2(E+M)\beta], \quad B = [\lambda - m^2 - 2(E+M)\alpha], \quad C = 0.$$

The remaining constants are thus calculated via (A5) as

$$c_4 = \frac{1}{4}, \quad c_5 = -\frac{1}{4}, \quad c_6 = \frac{1 + 4\lambda + 8(E+M)\beta}{4}, \quad c_7 = \frac{m^2 - \lambda + 2(E+M)\alpha}{4} - \frac{1}{8}, \quad c_8 = \frac{1}{16},$$

$$c_9 = \frac{m^2 + 2(E+M)(\alpha+\beta)}{4}, \quad c_{10} = \frac{1}{2}, \quad c_{11} = \sqrt{m^2 + 2(E+M)(\alpha+\beta)}, \quad c_{12} = \frac{1}{2},$$

$$c_{13} = \frac{1}{2}\sqrt{m^2 + 2(E+M)(\alpha+\beta)}. \tag{16}$$

We use the energy relation (A10) and the parametric coefficients given by Eqs. (15) and (16) to obtain a relationship between the separation constant $\lambda$ and the new non negative angular integer $\tilde{n}$ as

$$\lambda = l(l+1) = \left(2\tilde{n} + \tilde{m} + \frac{3}{2}\right)^2 - 2(E+M)\beta - \frac{1}{4}, \tag{17a}$$

$$l = \sqrt{\left(2\tilde{n} + \tilde{m} + \frac{3}{2}\right)^2 - 2(E+M)\beta} - \frac{1}{2}, \tag{17b}$$

$$\tilde{m} = \sqrt{m^2 + 2(E+M)(\alpha+\beta)}. \tag{17c}$$

Once the RS part of the potential (1) being disappeared after setting $\alpha = \beta = 0$ or simply letting the angular part $V_{RS}(\theta) = 0$, we can obtain $l = 2\tilde{n} + |m| + 1$, $m = 0, 1, 2, \cdots$. Hence, the angular part, $V_{RS}(\theta)$ has singularities at angles $\theta = P\pi$ ($P = 0, 1, 2, 3, \cdots$) as well as at very small and very large values of $r$.

Let us find the corresponding polar angular part of the wave function. We obtain the weight function via (A11) as

$$\rho(z) = z^{1/2}(1-z)^{\sqrt{m^2 + 2(E+M)(\alpha+\beta)}}, \tag{18}$$

which leads to the first part of the angular wave function through (A13) being expressed in terms of the Jacobi polynomial as

$$y_{\tilde{n}}(z) \sim P_{\tilde{n}}^{\left(1/2, \sqrt{m^2 + 2(E+M)(\alpha+\beta)}\right)}(1-2z). \tag{19}$$

The second part of the angular wave function can be obtained via (A12) as

$$\phi(z) \sim z^{1/2}(1-z)^{\sqrt{m^2 + 2(E+M)(\alpha+\beta)}/2}, \tag{20}$$

and therefore the angular part of the wave function can be obtained via Eq. (A14); namely, $\Theta_l(z) = \phi(z) y_{\tilde{n}}(z)$ as

$$\Theta_l(\theta) = A_{\tilde{n}} \cos\theta (\sin\theta)^{\sqrt{m^2 + 2(E+M)(\alpha+\beta)}} P_{\tilde{n}}^{\left(1/2, \sqrt{m^2 + 2(E+M)(\alpha+\beta)}\right)}(1 - 2\cos^2\theta), \quad (21)$$

where $A_{\tilde{n}}$ is the normalization factor. When the RS part of potential (1) is being set to zero, i.e., $\alpha = \beta = 0$, then

$$\Theta_l(\theta) = A_{\tilde{n}} \cos\theta \sin^{|m|}\theta P_{\tilde{n}}^{(1/2, |m|)}(1 - 2\cos^2\theta).$$

### 3.2. Solution of the radial part

We will consider the energy eigenvalue equation and the wave function of the radial part of the Dirac equation with Hulthén potential. The exact solution is not handy due to existence of the centrifugal term $\lambda / r^2$ in Eq. (12a). Therefore, an approximate analytical solution has been done as

$$\frac{\lambda}{r^2} \approx \lambda v^2 \left[ d_0 + \frac{e^{-vr}}{\left(1 - e^{-vr}\right)^2} \right], \quad v = \frac{1}{a}, \quad r \ll a, \quad (22)$$

where $d_0$ is dimensioless constant, $d_0 = 1/12$ [45,46]. With the approximation (22) and the change of variables $s = e^{-vr}$, the radial equation (12a) becomes

$$U''_{nl}(s) + \frac{1-s}{s-s^2} U'_{nl}(s) + \frac{1}{\left(s-s^2\right)^2} \left[ -(\sigma + \varepsilon) s^2 + (2\varepsilon + \sigma - \lambda)s - \varepsilon \right] U_{nl}(s) = 0, \quad (23)$$

with

$$-\varepsilon = a^2 \left( E^2 - M^2 \right) - l(l+1) d_0, \quad \sigma = 2a^2 (E+M) V_0. \quad (24)$$

Comparing Eq. (23) with its counterpart hypergeometric equation (A2), we identify values of the following constants:

$$c_1 = 1, \quad c_2 = 1, \quad c_3 = 1, \quad A = \sigma + \varepsilon, \quad B = 2\varepsilon + \sigma - \lambda, \quad C = \varepsilon, \quad (25)$$

and the remaining constants are calculated via (A5) as

$$c_4 = 0, \quad c_5 = -\frac{1}{2}, \quad c_6 = \frac{1}{4}[1 + 4(\sigma + \varepsilon)], \quad c_7 = -2\varepsilon - \sigma + \lambda, \quad c_8 = \varepsilon,$$

$$c_9 = \frac{1}{4}(2l+1)^2, \quad c_{10} = 2\sqrt{\varepsilon}, \quad c_{11} = 2l+1, \quad c_{12} = \sqrt{\varepsilon}, \quad c_{13} = l+1. \quad (26)$$

Hence, the energy eigenvalue equation can be obtained via the relation (A10) and values of constants given by Eqs. (24)-(26) after lengthy but straightforward manipulations as

$$E^2 - M^2 = \frac{l(l+1)d_0}{a^2} - \frac{1}{4a^2}\left[\frac{\sigma}{(n+l+1)} - (n+l+1)\right]^2,$$

or alternatively

$$E^2 - M^2 = \frac{d_0}{a^2}\left[\left(2\tilde{n} + \tilde{m} + \frac{3}{2}\right)^2 - 2(E+M)\beta - \frac{1}{4}\right] - \frac{1}{4a^2}$$

$$\times \left[\frac{2(M+E)V_0 a^2}{\left(n + \sqrt{\left(2\tilde{n} + \tilde{m} + \frac{3}{2}\right)^2 - 2(E+M)\beta} + \frac{1}{2}\right)} - \left(n + \sqrt{\left(2\tilde{n} + \tilde{m} + \frac{3}{2}\right)^2 - 2(E+M)\beta} + \frac{1}{2}\right)\right]^2, \quad (27)$$

where $\tilde{m}$ is given in (17c).

We need to solve Eq. (27) numerically taking the following values of the parameters for the Hulthén potential: $V_0 = 3.4\,fm^{-1}$, $\delta = a^{-1} = 0.25\,fm^{-1}$, $M = 5\,fm^{-1}$ [45]. As shown in Table 1, Eq. (27) admits two negatively attractive bound-state energy solutions corresponding to the particle and its antiparticle. We have also shown these two solutions by studying the Hulthén plus a new ring-shaped potential and the Hulthén potential cases. As seen the existence or absence of the RS potential has no much effects on the energy states of the Hulthén particles with a very narrow band spectrum. However, the existence of the RS potential has a strong effect on the energy spectrum of the antiparticles as they change from attractive to repulsive for some states with a very wide band spectrum. Further, both particle and antiparticle are trapped by the Hulthén field as the ring-shaped potential is existing or disappearing.

To sharpen our analysis, we have also drawn the behavior of the bound-state energy with the potential depth $V_0$ and the screening parameter $\delta$ for the Hulthén potential and the RS Hulthén potential as shown in Figures 2 to 5.

The variation of the energy with $V_0$ for the RS Hulthén potential and Hulthén potential plotted in Figure 2 and Figure 4, respectively, shows that energy becomes more negative (strongly attractive) as $V_0$ increasing. Both particle and antiparticle are trapped by the Hulthén field. On the other hand, the variation of the energy with $\delta$ for the RS Hulthén potential and Hulthén potential plotted in Figure 3 and Figure 5, respectively, indicates that energy becomes less negative (weakly attractive) as $\delta$ increasing.

Let us find the corresponding radial part of the wave function. We find the weight function via (A11) as

$$\rho(s) = s^{2\sqrt{\varepsilon}} (1-s)^{2l+1}, \tag{28}$$

Hence, with Eq. (28), the first part of the radial wave function can be obtained by means of the relation (A13) in terms of the Jacobi polynomials as

$$y_n(s) \sim P_n^{(2\sqrt{\varepsilon}, 2l+1)}(1-2s). \tag{29}$$

We have used the definition of the Jacobi polynomials given by [46]

$$P_n^{(a,b)}(y) = \frac{(-1)^n}{n! 2^2} (1-y)^{-a} (1+y)^{-b} \frac{d^n}{dy^n} \left[ (1-y)^{a+n} (1+y)^{b+n} \right]. \tag{30}$$

The second part of the radial wave function can be obtained via (A12) as

$$\phi(s) \sim s^{\sqrt{\varepsilon}} (1-s)^{l+1}, \tag{31}$$

and thus the radial part of the wave function, $U_{nl}(s) = \phi(s) y_n(s)$ is

$$U_{nl}(r) = A_{nl} \left( e^{-r/a} \right)^{\sqrt{\varepsilon}} \left( 1 - e^{-r/a} \right)^{l+1} P_n^{(2\sqrt{\varepsilon}, 2l+1)} \left( 1 - 2e^{-r/a} \right), \tag{32}$$

where the normalization constant $A_{nl}$ has been calculated in Appendix B.

Finally, combining Eqs. (14), (21) and (32), the total upper-component of the wave function (11) becomes

$$\varphi(\vec{r}) = N_{n\tilde{n}m} \frac{1}{\sqrt{2\pi}} \cos\theta (\sin\theta)^{\tilde{m}} P_{\tilde{n}}^{(1/2, \tilde{m})} \left( 1 - 2\cos^2\theta \right) e^{\pm im\phi}$$

$$\times \left( e^{-r/a} \right)^{\sqrt{\varepsilon}} \left( 1 - e^{-r/a} \right)^{l+1} P_n^{(2\sqrt{\varepsilon}, 2l+1)} \left( 1 - 2e^{-r/a} \right), \tag{33}$$

where $m = 0, 1, 2, \cdots$, $n = 0, 1, 2, \cdots$, and recalling that

$$l = \sqrt{\left( 2\tilde{n} + \tilde{m} + \frac{3}{2} \right)^2 - 2(E+M)\beta} - \frac{1}{2}, \quad l = 0, 1, 2, \cdots.$$

The lower-component of the wave function (5) can be found by means of Eq. (8b) as

$$\chi(\vec{r}) = N_{n\tilde{n}m} \frac{1}{\sqrt{2\pi}} \frac{\vec{\sigma} \cdot \vec{p}}{E+M} \cos\theta (\sin\theta)^{\tilde{m}} P_{\tilde{n}}^{(1/2, \tilde{m})} \left( 1 - 2\cos^2\theta \right) e^{\pm im\phi}$$

$$\times \left( e^{-r/a} \right)^{\sqrt{\varepsilon}} \left( 1 - e^{-r/a} \right)^{l+1} P_n^{(2\sqrt{\varepsilon}, 2l+1)} \left( 1 - 2e^{-r/a} \right), \quad E \neq -M \tag{34}$$

To avoid the repetition, its worthy to note that in the case of exact p-spin symmetry when $S(\vec{r}) = -V(\vec{r})$ or $\Sigma(\vec{r}) = S(\vec{r}) + V(\vec{r}) = 0$ and $\Delta(\vec{r}) = V(\vec{r}) - S(\vec{r}) = 2V(\vec{r})$, we found that it is necessary to perform the following mappings of Eqs. (8a) and (8b) as [47]

$$\phi(r) \to \chi(r), \ \chi(r) \to -\phi(r), \ V(r) \to -V(r), \ E \to -E, \tag{35}$$

so to obtain the solution of the present case.

## 4. The non-relativistic limit

Here, we consider the non-relativistic solutions for the new generalized ring-shaped Hulthén potential and the Hulthén potential:

### 4.1. A new generalized ring-shaped Hulthén potential case

In finding the non-relativistic solution, we let $E - M \approx E_{nl}$ and $E + M \approx 2\mu$. Hence, the bound state energy formula can be easily obtained via Eq. (27) as

$$E_{n\tilde{n}l} = \frac{1}{2\mu a^2} \left\{ \left( N^2 - 4\mu\beta - \frac{1}{4} \right) d_0 - \left[ \frac{2\mu V_0 a^2}{(n+l+1)} - \frac{(n+l+1)}{2} \right]^2 \right\}, \tag{36}$$

where $l$ and $N$ are given by

$$l = -\frac{1}{2} + \sqrt{N^2 - 4\mu\beta}, \ N = 2\tilde{n} + \tilde{m} + \frac{3}{2}, \ \tilde{m} = \sqrt{m^2 + 4\mu(\alpha+\beta)}. \tag{37}$$

where $m = 0,1,2,\cdots$ and $\tilde{n} = 0,1,2,\cdots$, Further, the non-relativistic wave function can be found by using Eqs. (24) and (33) as

$$\psi(r,\theta,\phi) = N_{n\tilde{n}l} \frac{1}{\sqrt{2\pi}} \cos\theta (\sin\theta)^{\tilde{m}} P_{\tilde{n}}^{(1/2,\tilde{m})}(1-2\cos^2\theta) e^{\pm im\phi}$$

$$\times \left(e^{-r/a}\right)^{\sqrt{\kappa}} \left(1-e^{-r/a}\right)^{l+1} P_n^{(2\sqrt{\kappa},2l+1)}(1-2e^{-r/a}), \tag{38}$$

with

$$\kappa = -2\mu a^2 E_{nl} + \frac{l(l+1)}{12}, \ E_{nl} < 0. \tag{39}$$

### 4.2. Hulthén potential case

Letting $\alpha = \beta = 0$, Eqs. (36) and (38) become

$$E_{nl} = \frac{1}{2\mu a^2} \left\{ l(l+1)d_0 - \left[ \frac{2\mu V_0 a^2}{(n+l+1)} - \frac{(n+l+1)}{2} \right]^2 \right\}, \tag{40}$$

which is identical to Ref. [54] and the wave function is written as

$$\psi(r,\phi) = N_{nl} \frac{1}{\sqrt{2\pi}} e^{\pm im\phi} \left(e^{-r/a}\right)^{\sqrt{\varepsilon}} \left(1-e^{-r/a}\right)^{l+1} P_n^{(2\sqrt{\varepsilon},2l+1)}\left(1-2e^{-r/a}\right), \tag{41}$$

where $\varepsilon$ is given in Eq. (39).

## 5. Concluding Remarks

In this work, we have investigated the approximate bound state solutions of the Dirac equation with the Hulthén plus a new RS potential for any orbital $l$ quantum numbers. By making an appropriate approximation to deal with the centrifugal term, we have obtained the energy eigenvalue equation and the normalized two spinor components of the wave function $\varphi(\vec{r})$ and $\chi(\vec{r})$ expressed in terms of the Jacobi polynomials. This problem is solved within the shortcut of NU method introduced recently in [46]. The relativistic solution can be reduced into the Schrödinger solution under the nonrelativistic limit.

**Acknowledgments**
We thank the kind referees for the positive enlightening comments and suggestions, which have greatly helped us in making improvements to this paper.

**Appendix A: A Shortcut of the NU Method**

The NU method is used to solve second order differential equations with an appropriate coordinate transformation $s = s(r)$ [24]

$$\psi_n''(s) + \frac{\tilde{\tau}(s)}{\sigma(s)}\psi_n'(s) + \frac{\tilde{\sigma}(s)}{\sigma^2(s)}\psi_n(s) = 0, \tag{A1}$$

where $\sigma(s)$ and $\tilde{\sigma}(s)$ are polynomials, at most of second degree, and $\tilde{\tau}(s)$ is a first-degree polynomial. To make the application of the NU method simpler and direct without need to check the validity of solution. We present a shortcut for the method. So, at first we write the general form of the Schrödinger-like equation (A1) in a more general form applicable to any potential as follows [47-50]

$$\psi_n''(s) + \frac{(c_1 - c_2 s)}{s(1 - c_3 s)}\psi_n'(s) + \frac{(-As^2 + Bs - C)}{s^2(1 - c_3 s)^2}\psi_n(s) = 0, \tag{A2}$$

satisfying the wave functions

$$\psi_n(s) = \phi(s) y_n(s). \tag{A3}$$

Comparing (A2) with its counterpart (A1), we obtain the following identifications:

$$\tilde{\tau}(s) = c_1 - c_2 s, \quad \sigma(s) = s(1 - c_3 s), \quad \tilde{\sigma}(s) = -As^2 + Bs - C, \tag{A4}$$

Following the NU method [24], we obtain the following necessary parameters [48],

(i) Relevant constant:

$$c_4 = \frac{1}{2}(1 - c_1), \qquad c_5 = \frac{1}{2}(c_2 - 2c_3),$$

$$c_6 = c_5^2 + A, \qquad c_7 = 2c_4 c_5 - B,$$

$$c_8 = c_4^2 + C, \qquad c_9 = c_3(c_7 + c_3 c_8) + c_6,$$

$$c_{10} = c_1 + 2c_4 + 2\sqrt{c_8} - 1 > -1, \qquad c_{11} = 1 - c_1 - 2c_4 + \frac{2}{c_3}\sqrt{c_9} > -1, \; c_3 \neq 0,$$

$$c_{12} = c_4 + \sqrt{c_8} > 0, \qquad c_{13} = -c_4 + \frac{1}{c_3}(\sqrt{c_9} - c_5) > 0, \; c_3 \neq 0. \tag{A5}$$

(ii) Essential polynomial functions:

$$\pi(s) = c_4 + c_5 s - \left[\left(\sqrt{c_9} + c_3\sqrt{c_8}\right)s - \sqrt{c_8}\right], \tag{A6}$$

$$k = -(c_7 + 2c_3 c_8) - 2\sqrt{c_8 c_9}, \tag{A7}$$

$$\tau(s) = c_1 + 2c_4 - (c_2 - 2c_5)s - 2\left[\left(\sqrt{c_9} + c_3\sqrt{c_8}\right)s - \sqrt{c_8}\right], \tag{A8}$$

$$\tau'(s) = -2c_3 - 2\left(\sqrt{c_9} + c_3\sqrt{c_8}\right) < 0. \tag{A9}$$

(iii) Energy equation:

$$c_2 n - (2n+1)c_5 + (2n+1)\left(\sqrt{c_9} + c_3\sqrt{c_8}\right) + n(n-1)c_3 + c_7 + 2c_3 c_8 + 2\sqrt{c_8 c_9} = 0. \tag{A10}$$

(iv) Wave functions

$$\rho(s) = s^{c_{10}}(1-c_3 s)^{c_{11}}, \tag{A11}$$

$$\phi(s) = s^{c_{12}}(1-c_3 s)^{c_{13}}, \quad c_{12} > 0, \quad c_{13} > 0, \tag{A12}$$

$$y_n(s) = P_n^{(c_{10}, c_{11})}(1 - 2c_3 s), \quad c_{10} > -1, \quad c_{11} > -1, \tag{A13}$$

$$\psi_{n\kappa}(s) = N_{n\kappa} s^{c_{12}}(1-c_3 s)^{c_{13}} P_n^{(c_{10}, c_{11})}(1 - 2c_3 s). \tag{A14}$$

where $P_n^{(\mu,\nu)}(x)$, $\mu > -1$, $\nu > -1$, and $x \in [-1,1]$ are Jacobi polynomials with

$$P_n^{(a_0, b_0)}(1 - 2s) = \frac{(a_0 + 1)_n}{n!} {}_2F_1(-n, 1 + a_0 + b_0 + n; a_0 + 1; s), \tag{A15}$$

and $N_{n\kappa}$ is a normalization constant. Also, the above wave functions can be expressed in terms of the hypergeometric function as

$$\psi_{n\kappa}(s) = N_{n\kappa} s^{c_{12}}(1-c_3 s)^{c_{13}} {}_2F_1(-n, 1 + c_{10} + c_{11} + n; c_{10} + 1; c_3 s) \tag{A16}$$

where $c_{12} > 0$, $c_{13} > 0$ and $s \in [0, 1/c_3]$, $c_3 \neq 0$.

**Appendix B: Calculation of the Normalization Constant**

To compute the normalization constant $A_{nl}$ in Eq. (32), it is easy to show with the use of $R_{nl}(r) = r^{-1} U_{nl}(r)$, that

$$\int_0^\infty |R_{nl}(r)|^2 r^2 dr = \int_0^\infty |U_{nl}(r)|^2 dr = \int_0^1 |U_{nl}(s)|^2 \frac{ads}{s} = 1, \tag{B1}$$

where we have used the substitution $s = e^{-r/a}$. Inserting Eq. (32) into (B1) and using the following definition of the Jacobi polynomial [51]

$$P_n^{(p_0, w_0)}(1 - 2s) = \frac{\Gamma(n + p_0 + 1)}{n! \Gamma(p_0 + 1)} {}_2F_1(-n, p_0 + w_0 + n + 1; 1 + p_0; s), \tag{B2}$$

where $p_0 = 2\sqrt{\varepsilon}$ and $w_0 = 2l + 1$, we arrive at

$$|A_{nl}|^2 \int_0^1 s^{2\sqrt{\varepsilon}-1}(1-s)^{2l+2} \left\{ {}_2F_1\left(-n, 2\sqrt{\varepsilon} + 2l + 2 + n; 1 + 2\sqrt{\varepsilon}; s\right) \right\}^2 ds$$

$$= \frac{1}{a} \left( \frac{n!\Gamma(2\sqrt{\varepsilon}+1)}{\Gamma(n+2\sqrt{\varepsilon}+1)} \right)^2. \tag{B3}$$

where $_2F_1$ is the hypergeometric function. Using, the following integral formula [52,53]

$$\int_0^1 z^{2\lambda-1}(1-z)^{2(\eta+1)} \left\{ _2F_1(-n, 2(\lambda+\eta+1)+n; 1+2\lambda+1; z) \right\}^2 dz$$

$$= \frac{(n+\eta+1)n!\Gamma(n+2\eta+2)\Gamma(2\lambda)\Gamma(2\lambda+1)}{(n+\eta+\lambda+1)\Gamma(n+2\lambda+1)\Gamma(n+2\lambda++2\eta+2)}, \quad \eta > -\frac{3}{2}, \ \lambda > 0, \tag{B4}$$

we can get the normalization constant as

$$A_{nl} = \sqrt{\frac{2\sqrt{\varepsilon}n!(n+l+\sqrt{\varepsilon}+1)\Gamma(n+2l+2\sqrt{\varepsilon}+2)}{a(n+l+1)\Gamma(n+2l+2)\Gamma(n+2\sqrt{\varepsilon}+1)}}. \tag{B5}$$

The relation (B5) can be used to compute the normalization constant for $n = 0, 1, 2, \cdots$. for the ground state ($n = 0$), we have

$$A_{0l} = \sqrt{\frac{(l+\sqrt{\varepsilon}+1)}{a(l+1)B(2\sqrt{\varepsilon}, 2l+2)}}, \tag{B6}$$

where

$$B(2\sqrt{\varepsilon}, 2l+2) = \frac{\Gamma(2l+2)\Gamma(1+2\sqrt{\varepsilon})}{2\sqrt{\varepsilon}\Gamma(2l+2+2\sqrt{\varepsilon})}. \tag{B7}$$

**Table 1.** The bound state energy eigenvalues of the Dirac equation in units of $fm^{-1}$ with the ring-shaped-Hulthén potential and the Hulthén potential cases.

| $n$ | $\tilde{n}$ | $m$ | $\alpha = \beta = 1$ | | $\alpha = \beta = 0$ | |
|---|---|---|---|---|---|---|
| | | | $E^1_{n,\tilde{n},m}$ | $E^2_{n,\tilde{n},m}$ | $E^1_{n,\tilde{n},m}$ | $E^2_{n,\tilde{n},m}$ |
| 0 | 0 | 0 | −4.995583758 | −4.139490168 | −4.996058414 | −4.720283669 |
| 1 | 0 | 0 | −4.988883706 | −3.351144541 | −4.990126164 | −4.388096150 |
| 1 | 0 | 1 | −4.983219226 | −3.181185632 | −4.983417636 | −3.949703384 |
| 1 | 1 | 0 | −4.972069376 | −1.450852722 | −4.975125259 | −3.429332733 |
| 1 | 1 | 1 | −4.964591276 | −1.33435789 | −4.965242671 | −2.851556752 |
| 2 | 0 | 0 | −4.979466843 | −2.494855449 | −4.981810530 | −3.951310489 |
| 2 | 0 | 1 | −4.972286615 | −2.346989231 | −4.972714353 | −3.431743639 |
| 2 | 1 | 0 | −4.957273060 | −0.642877021 | −4.962026060 | −2.854773363 |
| 2 | 1 | 1 | −4.948594597 | −0.544726664 | −4.949737250 | −2.243800108 |
| 2 | 2 | 0 | −4.927892096 | 0.866047610 | −4.935838511 | −1.619431412 |
| 2 | 2 | 1 | −4.918016190 | 0.933230388 | −4.920319274 | −0.998615904 |
| 3 | 0 | 0 | −4.967330159 | −1.644145559 | −4.971102883 | −3.433355109 |
| 3 | 0 | 1 | −4.958826910 | −1.517571480 | −4.959607143 | −2.857192280 |
| 3 | 1 | 0 | −4.958826910 | −1.517571480 | −4.946508429 | −2.247028930 |
| 3 | 1 | 1 | −4.92998072 | 0.1758391135 | −4.931796375 | −1.623473548 |

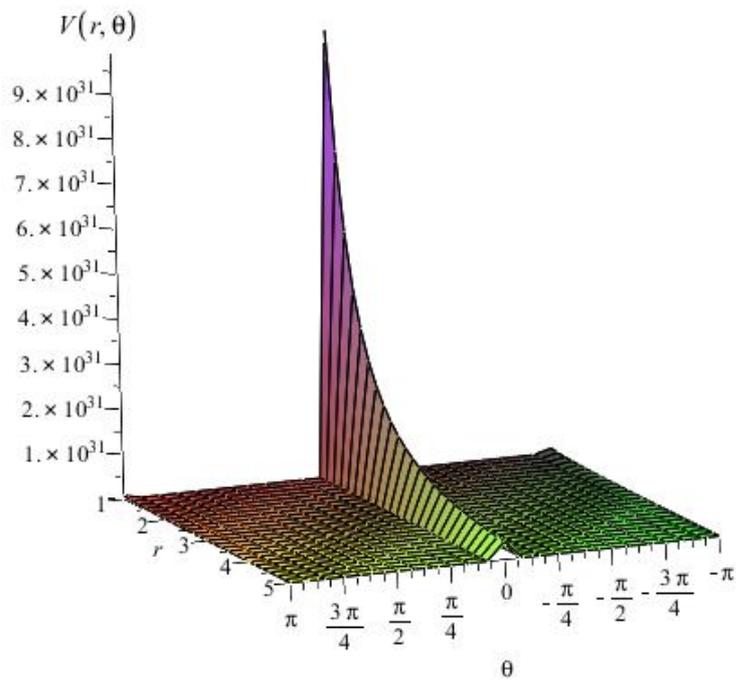

**Fig. 1:** A plot of ring-shaped Hulthén potential.

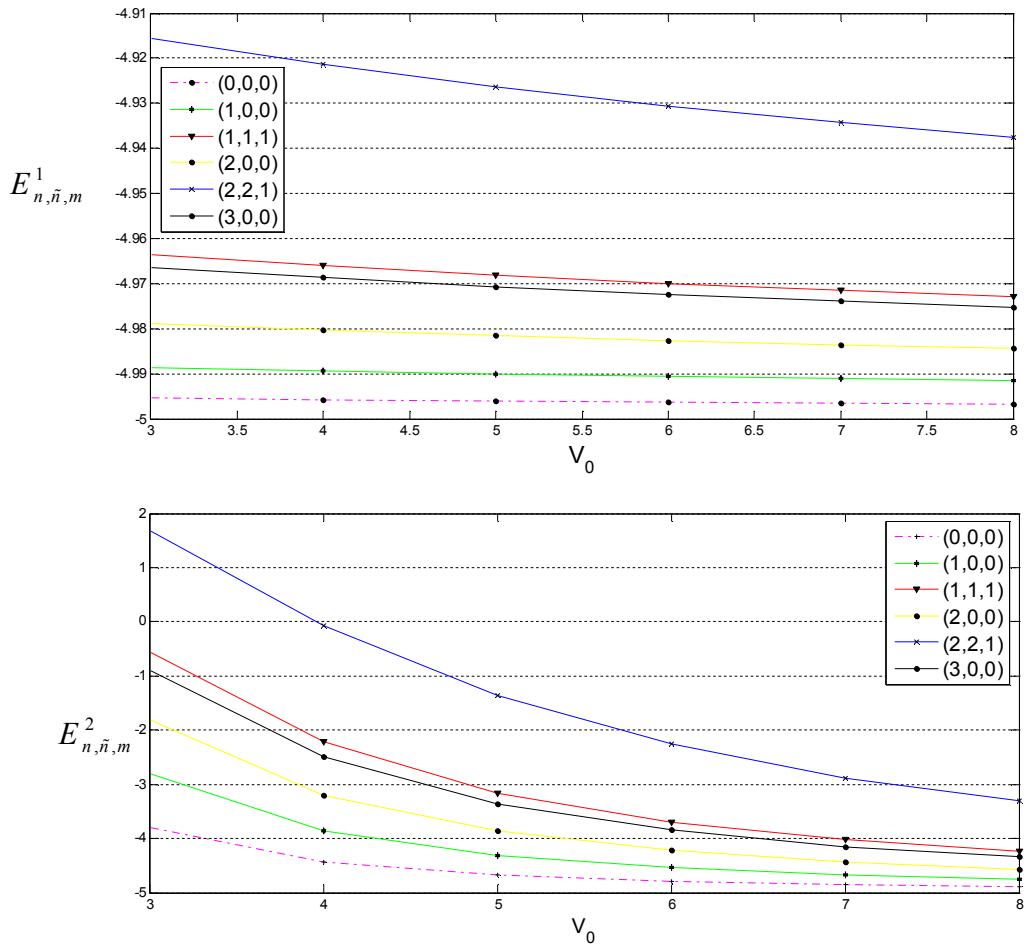

**Fig. 2:** Energy behavior of the Dirac equation with Hulthén potential plus RS potential versus potential depth $V_0$ for various $n$, $\tilde{n}$ and $m$, respectively

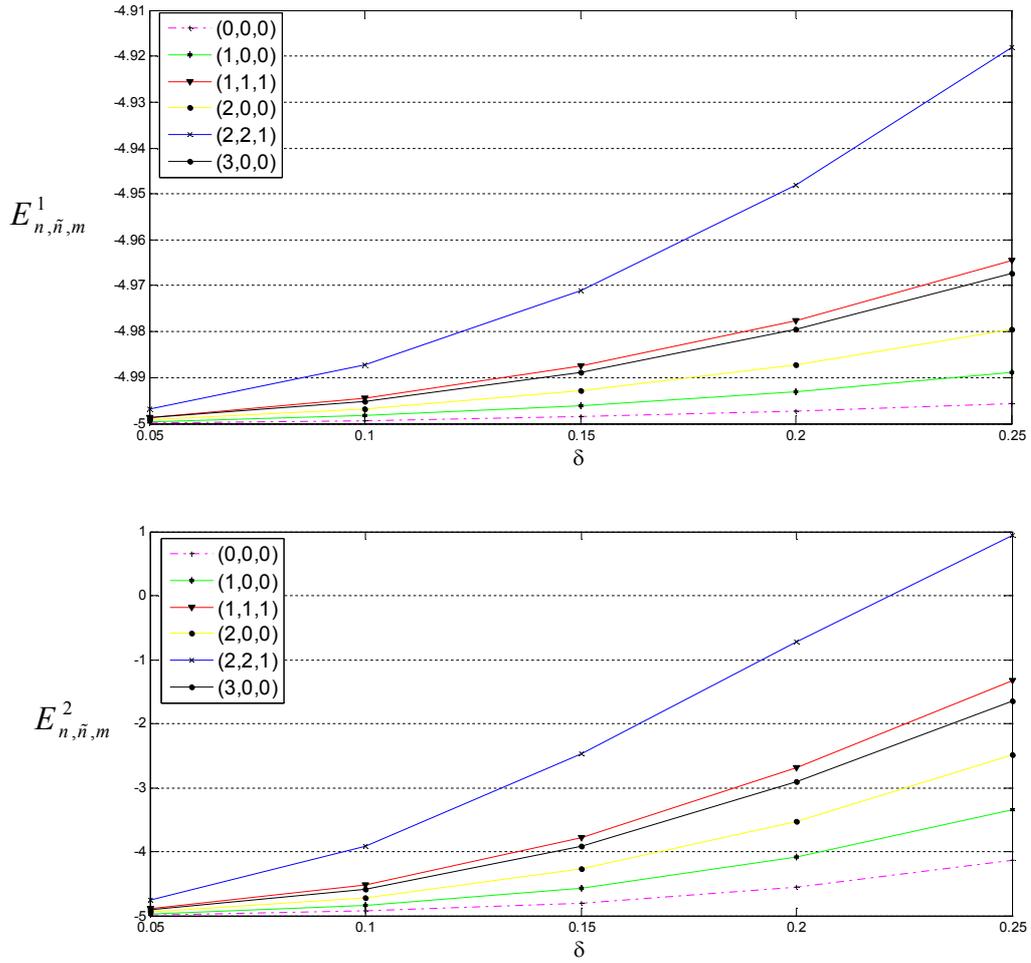

**Fig. 3:** Energy behavior of the Dirac equation with Hulthén potential plus RS potential versus screening parameter $\delta = a^{-1}$ for various $n$, $\tilde{n}$ and $m$, respectively

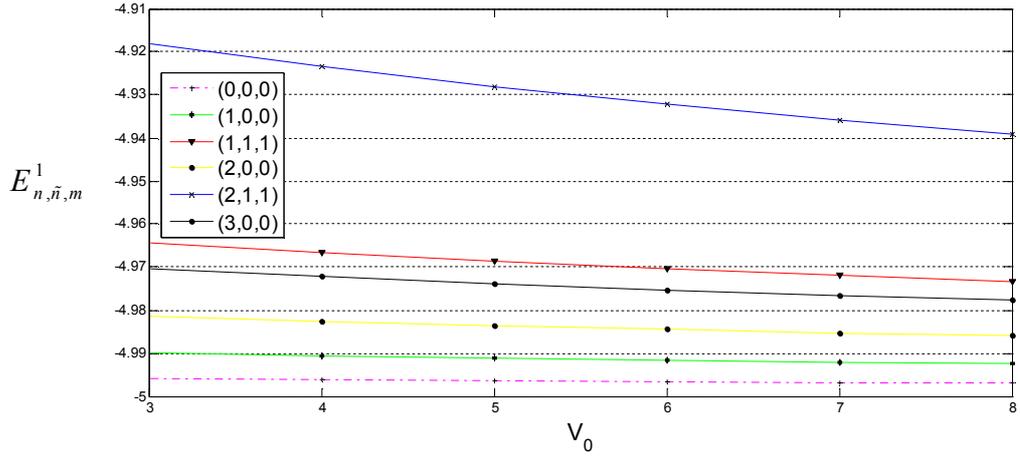

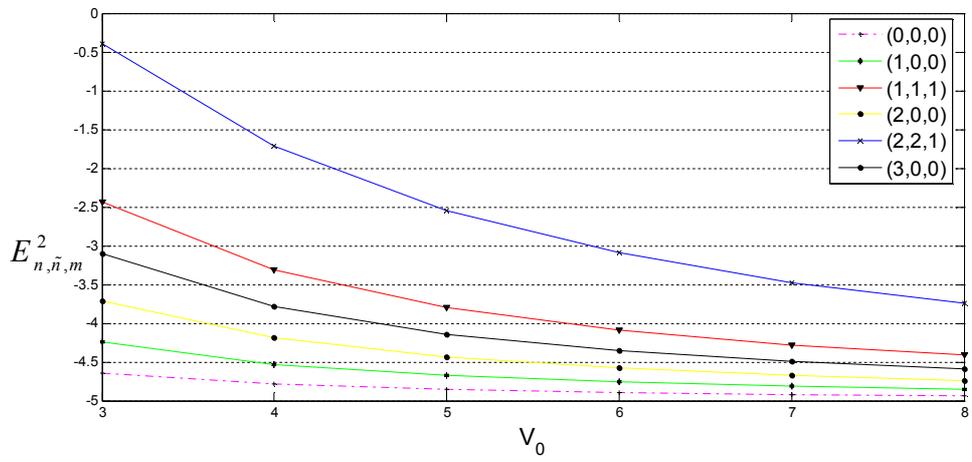

**Fig. 4:** Energy behavior of the Dirac equation with Hulthén potential without RS potential ($\alpha = \beta = 0$) versus potential depth $V_0$ for various $n$, $\tilde{n}$ and $m$, respectively

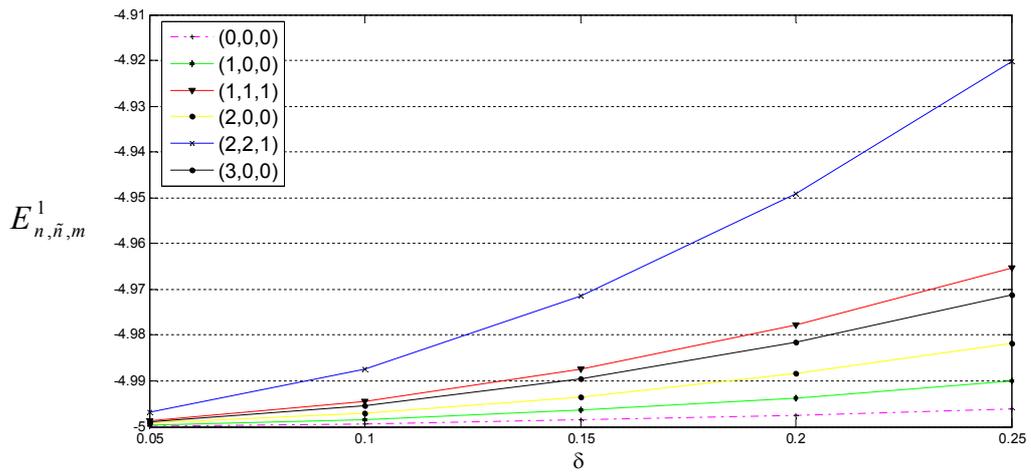

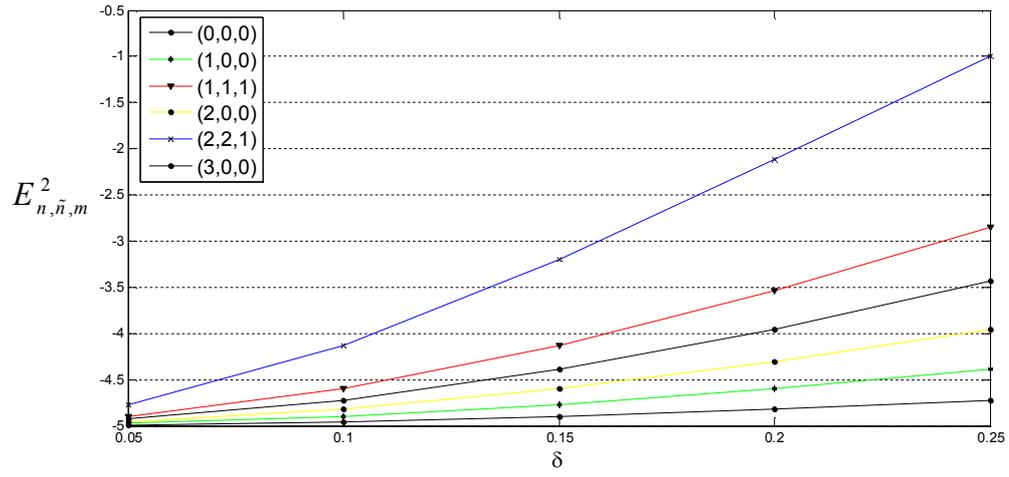

**Fig. 5:** Energy behavior of the Dirac equation with Hulthén potential without RS potential ($\alpha = \beta = 0$) versus screening parameter $\delta = a^{-1}$ for various $n$, $\tilde{n}$ and $m$, respectively